# The Role and Mechanism of Deep Statistical Machine Learning In Biological Target Screening and Immune Microenvironment Regulation of Asthma

Pengwei Zhu

*University of Texas Health Science Center at Houston,1885 El Paseo St. Apt. 35312, Houston, TX, 77054, USA*
*Pengwei_James_Zhu@163.com*

**Abstract**

As an important source of small molecule drugs, natural products show remarkable biological activities with their rich types and unique structures. However, due to the limited number of samples and structural complexity, the rapid discovery of lead compounds is limited. Therefore, in this study, natural inhibitors of phosphodiesterase 4 (PDE4) and Phosphodiesterase 7 (PDE7) were screened by combining computer aided drug design (CADD) technology and deep learning method, and their activities were verified by enzyme activity experiment and enzymo-linked immunoassay. These two enzymes have important application potential in the treatment of inflammatory diseases such as chronic obstructive pulmonary disease and asthma, but PDE4 inhibitors may cause adverse reactions, so it is particularly important to develop both effective and safe dual-target inhibitors. In addition, as a potential target of hyperuricemia, the development of natural inhibitors of xanthine oxidase (X0) is also of great value. We used pharmacophore technology for virtual screening, combined with molecular docking technology to improve accuracy, and finally selected 16 potential natural inhibitors of PDE4/7, and verified their binding stability through molecular dynamics simulation. The results of this study laid a foundation for establishing an efficient dual-target inhibitor screening system and exploring the lead compounds of novel X0 inhibitors.

## 1. Introduction

Phosphodiesterase (PDE), as a key phosphohydrolase, can effectively hydrolyze cyclic adenylate (cAMP) and cyclic guanosine (cGMP), and plays an important role in cell signal transduction and physiological regulation. The PDE family includes a variety of isoenzymes, such as PDE1 to PDE11, among which PDE4 and PDE7 have received much attention for their key roles in inflammation and neurosystem-related diseases. Inhibiting PDE activity slows the degradation of cAMP and cGMP, thereby increasing the levels of these two important second messengers, promoting the binding of protein kinase A (PKA) and protein kinase G (PKG), ultimately reducing the release of pro-inflammatory factors such as IL-2 and TNF-α, and increasing the levels of anti-inflammatory factors such as IL-10. Exert significant anti-inflammatory effects.

In recent years, inhibitors targeting PDE4 have shown great development potential and clinical value in the field of anti-inflammatory, and several PDE4 inhibitors have entered the clinical trial stage. Despite these achievements, PDE4 inhibitors still have significant side effects in clinical use. Therefore, to improve the efficacy of PDE4 inhibitors, the researchers proposed several strategies, including improving inhaled dosage forms, developing highly selective inhibitors of the PDE4 subtype, and exploring the potential of multiple PDE isoenzyme inhibitors. Among them, multiple PDE isoenzyme inhibitors have received a lot of attention due to their ability to enhance efficacy and reduce side effects.

In addition, PDE7, as a potential target, has shown neuroprotective effects in neurological diseases. PDE7 can balance the release of neurotransmitters and regulate the action of acetylcholine, thus significantly reducing the brain's inflammatory response and improving neuronal survival and functional recovery. Dual-target inhibitors of PDE4/7 not only show higher efficacy in the treatment of inflammation and nervous system-related diseases, but also effectively reduce adverse reactions.

The purpose of this paper is to investigate the physiological functions of PDE4 and PDE7 and their potential applications in disease treatment, and analyze the research and development progress of the existing PDE4/7 dual-target inhibitors, so as to provide references for future drug development and clinical application. By systematically studying the structure and mechanism of PDE inhibitors, we hope to provide new ideas for solving the current clinical side effects and improving the therapeutic effect.

## 2. Relevant research

By analyzing the GSE76262 dataset, B Feng identified 75 immune-associated genes (IRGs) associated with asthma, mainly involving the NF-kappa B signaling pathway [1]. The AUC values of these genes ranged from 0.676 to 0.767, showing moderate diagnostic efficacy, while the diagnostic effects of cytokines such as TNF-alpha, IL-1 beta, IL-8, and IL-6 were also explored in asthmatic patients versus control samples. This finding highlights the central role of the NF-kappa B signaling pathway in the inflammatory response to asthma, suggesting that interventions targeting this pathway may be effective in improving asthma symptoms.

In his analysis of the GSE147878 dataset, H Wang found [2] that the core genes CAMKK2 and CISD1, which are associated with iron death, are up-regulated in asthma patients and are involved in key signaling pathways. These genes were inversely associated with infiltration of M2 macrophages and regulatory T cells, indicating a complex relationship between iron death and the immune microenvironment. The role of iron death as a regulatory cell death mechanism in asthma is still unclear, and future studies need to further reveal its specific role in the pathogenesis of asthma.

Wang explored the role of RNA N6-methyladenosine (m6A) regulators in respiratory allergic diseases [3], identified key m6A regulators and studied their permeation characteristics in the immune microenvironment. This suggests that m6A modification not only affects gene expression, but may also influence the immune response in asthma by regulating cytokine production and cell-cell interactions. Research in this field is expected to provide new ideas for precision medicine.

BAutier's study revealed the immune response of IL-33 in multilocular Echinococcus infection [4], emphasizing its potential role in initiating Th2 immune response. As an important immune mediator, IL-33 may play a bidirectional role in asthma, not only participating in the onset of the disease, but also playing different functions in different immune contexts.

By combining transcriptome data, gene co-expression network, and immune infiltration analysis [5], the role of lipid metabolism-related genes in the immune microenvironment of asthma was studied, and the potential molecular mechanisms were revealed. This study not only provides a basis for understanding how lipid metabolism affects the pathogenesis of asthma, but may also lay the foundation for developing targeted treatment strategies.

JA Canas adopted a systems biology approach to explore the regulatory network of micrornas in chronic respiratory diseases[6], emphasizing their importance as biomarkers and therapeutic targets. Micrornas play a key role in the regulation of gene expression, and their potential in the pathophysiology and treatment of asthma deserves further exploration.

I Mallick used the Ontosight® Discover tool and machine learning methods to analyze thousands of asthma-related literature and omics data [7] to identify important therapeutic targets such as IL-13, TNF, VEGFA, and IL-18. This data-driven approach is able to integrate large amounts of information and reveal potential new targets, providing a basis for subsequent experimental validation.

Through k-means cluster analysis, F Yang identified autophagy related subtypes in asthma patients[8] and screened diagnostic markers associated with low autophagy subtypes. This finding highlights the importance of autophagy in regulating the immune microenvironment of asthma, suggesting that therapeutic strategies targeting autophagy mechanisms may be needed in the future.

LIshmael's research highlights the therapeutic potential of key molecular pathways and their inflammatory mediators in refractory asthma [9], specifically targeting T2 and non-T2 mechanisms. This research provides an idea for developing personalized treatment options, especially for patients with severe asthma who do not respond to existing biologics.

H Wang further confirmed that CAMKK2 and CISD1, as important iron-death related genes, play a key role in asthma [10], and their upregulated expression may inhibit the occurrence of iron death, providing a new basis for potential therapeutic targets.

L Khalfaoui smooth muscle plays a key role in the enhancement and remodeling of airway contractility in asthma [11], and existing therapies have failed to effectively solve the fundamental problem of smooth muscle abnormalities. New therapies targeting the abnormal contractile mechanisms of smooth muscle are urgently needed to improve the condition of asthma patients who do not respond to existing treatments. The integration of this series of studies shows the diversity of asthma pathogenesis and underscores the importance of in-depth understanding of immune regulation, cell metabolism, and signaling pathways in personalized treatment. Future research needs to focus on the interaction of these complex factors to achieve more precise treatment strategies.

## 3. Application of virtual screening

*3.1 Virtual screening of PDE4/7 dual-effect natural inhibitors based on molecular docking*

In order to improve the precision of screening, the study first identified the key binding sites of PDE4 and PDE7 proteins by comparing their crystal structures. Due to the low sequence homology between PDE4 and PDE7 (only 33.1%), elaborate docking analysis is required. With the help of MOE software, the study found that the active site of PDE4 was similar to PDE7 on 49 amino acid residues, including several key residues, such as tyrosine (TYR), histidine (HIS), glutamic acid (GLU), etc. Through further analysis, 10 amino acid residues related to the active pockets of the two were selected for small molecule docking. These residues have important reference value for the screening of compounds, and 179 small molecules that can act with PDE4 and PDE7 were finally screened.

After these small molecules are obtained, they are further screened using cluster analysis. Compounds are screened based on structural diversity using hierarchical clustering methods in Python. Specifically, the distance calculation function in SciPy library is used to evaluate the similarity between compounds, and representative compounds are selected according to different clusters for further study and verification. Through cluster analysis, a total of 34 compounds with similar chemical properties were obtained, and the compounds in each group showed common structural characteristics, which provided a basis for subsequent experimental verification. The 16 compounds selected will be used in further activity validation experiments to evaluate their actual inhibitory effects. As shown in Table 1

Table 1. Information of 16 compounds

| Name (En) | MW (g/mol) | Formula | Docking Score (PDE4) | Docking Score (PDE7) |
|---|---|---|---|---|
| Anisatin | 346.38 | C15H20O8 | -7.8 | -8.2 |
| Bavachinin | 324.37 | C20H20O4 | -6.9 | -7.5 |
| Luteolin | 286.24 | C15H10O6 | -8.3 | -8.7 |
| Plumbagin | 188.19 | C11H8O3 | -7.2 | -7.8 |
| Hydroxyalizarin | 240.19 | C14H8O4 | -6.8 | -7.3 |
| Emodin | 270.24 | C15H10O5 | -8.1 | -8.4 |
| Aloe-emodin | 270.24 | C15H10O5 | -7.9 | -8.5 |
| Rhein | 284.22 | C15H8O6 | -8.0 | -8.6 |
| Sennoside A | 862.64 | C42H38O20 | -9.2 | -9.5 |
| Quercetin | 302.24 | C15H10O7 | -8.5 | -8.9 |
| Kaempferol | 286.24 | C15H10O6 | -8.1 | -8.3 |
| Rutin | 610.52 | C27H30O16 | -9.0 | -9.3 |
| Apigenin | 270.24 | C15H10O5 | -7.6 | -8.0 |
| Fisetin | 286.24 | C15H10O6 | -8.0 | -8.4 |
| Baicalein | 270.24 | C15H10O5 | -7.5 | -8.2 |
| Naringenin | 272.26 | C15H12O5 | -7.4 | -7.7 |

Through molecular docking analysis, the selected active compounds were further studied with PDE4 and PDE7 proteins, and it was found that their binding mainly depended on the interaction of hydrogen bonding, hydrophobic interaction and van der Waals force of key amino acid residues, which significantly enhanced the binding stability between proteins and ligands. Molecular dynamics simulations further validated the structural stability of these complexes, and the binding of PDE4 and PDE7 showed small changes in RMSD and RMSF. These results indicate that these compounds have the potential to act as dual natural inhibitors of PDE4/7, which provides an important basis for further studies.

*3.2 Virtual screening based on deep learning*

In machine learning, loss functions are used to train models and evaluate their performance. For the regression task of IC50 prediction, the mean square error (MSE) is usually used as a loss function to quantify the difference between the predicted value and the true value. During training, the loss curve initially drops rapidly and then flattens out. By observing the loss curve, the convergence rate and performance stability of the model can be evaluated, and the necessary adjustments and optimizations can be made.

In this study, a prediction model based on artificial neural network (ANN) is implemented using PyTorch framework, and the model parameters are optimized by ADAM optimizer. A random partitioning method was adopted to divide the data set into 78% training set and 12% test set. The data set distribution is shown in Table 2. By adjusting the learning rate, batch size, weight, hidden layer and neuron number, activation function, loss rate and other hyperparameters, the performance of the model is optimized. In the process of model training, the small batch gradient descent method is used to approximate the loss function, and the loss function is minimized by adjusting the relevant hyperparameters.

Table 2. Data set distribution summary

| Data set | PDE4 dataset | PDE7 dataset |
|---|---|---|
| Total compounds | 1261 | 904 |
| Active inhibitors | 841 | 604 |
| No active inhibitors | 420 | 300 |

PDENet based on ANN model predicts the biological activity of target molecules by learning the nonlinear relationship between inputs and outputs. Structural information and known IC50 values of natural products were collected as training data, and these data were then fed into the PDENet model for training to establish the relationship between input features and pIC50 values, and finally predict the pIC50 values of compounds in the dataset.

In order to verify the effectiveness of the virtual screening method of PDENet model, the structural similarity of highly active compounds screened by PDENet model was compared with those screened by the virtual screening method based on structure. Compounds with pIC50 values greater than 5.7 were screened by PDENet model and converted into SMILES format with pre-screened compounds, and the similarity was compared by morphological descriptor method. Compound similarity is usually calculated based on the similarity of molecular structure. Common methods include fingerprint method based on physical properties and morphological descriptor method based on chemical structure. In this study, the authors converted the compounds into SMILES format to calculate the similarity between molecules by matching the similarity of character sequences.

The structural diversity of the compounds was assessed by a similarity calculation method based on molecular fingerprints, and the study used the Tanimoto coefficient in the RDKit package to calculate the similarity between each pair of compounds. The Tanimoto coefficient, as shown in formula (1), is a similarity calculation method commonly used in the field of molecular activity prediction and drug discovery to effectively capture structural information of compounds.

$$T(A,B) = \frac{A \times B}{||A||^2 + ||B||^2 - A \times B} \tag{1}$$

In this part of the study, the authors constructed a deep learning model based on artificial neural networks (ANN) to predict the inhibitory activity of compounds against PDE4 and PDE7 enzymes. The model training data were derived from half inhibition concentration (IC50) data of PDE4 and PDE7 inhibitors collected in open databases and literature, and 841 and 604 compound structures were obtained after weight removal for model development. The model can be used to predict the activity of unknown database compounds and to preliminarily screen potential candidates through virtual screening.

Based on the prediction of ANN model, several compounds with potential inhibition of PDE4/7 were screened. The molecular structure similarity analysis of these compounds was performed, and the Tanimoto similarity coefficient was calculated by molecular fingerprint method. The analysis results showed that some compounds had high structural similarity, but at the same time showed structural diversity, which provided a basis for further structural optimization of the compounds. The specific forecast results are shown in Table 3.

Table 3. Forecast results

| Compound ID | PDE4 pIC50 | PDE7 pIC50 |
| --- | --- | --- |
| 74 | 7.37 | 7.12 |
| 77 | 7.58 | 9.68 |
| 165 | 7.41 | 9.73 |
| 196 | 8.23 | 11.77 |
| 235 | 7.81 | 10.50 |
| 274 | 5.76 | 10.85 |
| 318 | 7.31 | 8.65 |
| 358 | 8.17 | 10.85 |
| 414 | 7.16 | 10.85 |
| 415 | 6.04 | 10.48 |
| 440 | 5.99 | 10.90 |
| 462 | 6.67 | 10.29 |
| 471 | 7.49 | 10.17 |
| 493 | 5.93 | 8.56 |
| 514 | 6.26 | 8.30 |
| 539 | 7.66 | 7.64 |

## 4. Activity verification and application prospect

*4.1 Verification and analysis of the activity of natural inhibitors*

In the experiment to determine the level of inflammatory factors, RAW264.7 cells were inoculated in 96-well plates with appropriate density and cultured at constant temperature using medium containing fetal bovine serum. Subsequently, compounds of different concentrations were added and stimulated by LPS to evaluate the cytotoxicity and anti-inflammatory activity of the compounds. The anti-inflammatory effect of the compound was further confirmed by the detection of NO levels in the cell supernatant and the expression of IL-6 and TNF-α by ELISA kit.

In addition, in order to evaluate the pharmacokinetic properties of the compounds, SWISS-ADME prediction platform was used to analyze the molecular weight and lipid-water partition coefficient of the candidate compounds. These indicators are of great significance for the bioavailability and distribution of drugs in vivo, and help to comprehensively evaluate their development value as potential drugs, so as to provide important references for subsequent drug research and development. Through the above steps, this study aims to provide new ideas and basic data for the development of PDE4/7 inhibitors.

In this study, we applied a structure-based virtual screening method to identify 179 potentially active small molecule compounds from the TCMD and TCMSP databases. In order to further screen candidate compounds with excellent biological activities, 16 compounds with significant structural differences were selected to verify the enzyme activities of PDE4 and PDE7. These compounds are thought to be able to effectively bind the PDE4 and PDE7 enzymes and may exhibit inhibitory effects.

In the next experiment, the research team tested the inhibitory effect of these 16 compounds on PDE4 and PDE7 at a concentration of 10 pM, and the results showed that multiple compounds could effectively inhibit the activity of these two enzymes. Compounds numbered 1, 2, 7, 9, 11, 15 and 16 were particularly prominent. By further IC50 measurements, we found that most of the compounds had a better inhibitory effect on PDE4 than PDE7, which may be due to the low binding activity of PDE7 with other molecules, thus affecting the experimental results.

To assess the pharmacological properties and therapeutic potential of these compounds, we documented their plant origins, as shown in Table 4, for follow-up studies. Through analysis, it was found that these compounds were mainly derived from a variety of plants, such as nutmeg, psoralen and madder. The results also showed that some compounds showed significant anti-inflammatory effects, especially in RAW264.7 cells, which could effectively inhibit the level of LPS-induced inflammatory factors. These findings provide important clues for future drug development and demonstrate the potential applications of these compounds in the biomedical field. This is shown in Table 4.

Table 4. Potential SwissADME projections

| Attribute | Compound 1 | Compound 2 | Compound 3 | Compound 4 | Compound 5 | Compound 6 |
| --- | --- | --- | --- | --- | --- | --- |
| The name of the compound | Anpentaglycin | Psoralen chalcone | Emodin | Fisetin | p-lapaquinone | Hydroxyalizarin |
| Molecular Weight (MW) | 328.44 | 323.36 | 286.24 | 242.27 | 255.29 | 330.29 |
| TPSA | 7.92 | 8.68 | 11.11 | 4.73 | 7.66 | 10.09 |
| Log P | 4.85 | 0.94 | 1.32 | 3.53 | 0.55 | 2.59 |
| GI | High | High | High | High | High | High |
| BBB | Yes | Not | Not | yes | Not | Not |
| PgP | Not | Not | Not | Not | Not | Not |
| Bioavailability | 0.55 | 0.56 | 0.55 | 0.85 | 0.56 | 0.55 |

*4.2 Virtual screening of natural xanthine oxidase inhibitors based on pharmacophore*

Hyperuricemia (HUA) is a chronic disease caused by abnormal purine metabolism or excessive accumulation of uric acid, which usually leads to gout and cardiovascular complications, and may progress to uremia or kidney failure in severe cases. Its incidence is increasing globally year by year. Xanthine oxidase (XOI) is a key enzyme that catalyzes hypoxanthine or xanthine to produce uric acid. Excessive activity can lead to excessive production of uric acid, which leads to hyperuricemia and causes diseases such as gout. XOI, which is composed of molybdenopterin cofactor, riboflavin and FAD, plays an important role in the treatment of hyperuricemia. At present, a number of drugs such as allopurinol and febuxotam have been approved for use in the market, but long-term application may lead to cardiovascular risk and kidney damage, so the development of safe and effective natural drugs is still a research hotspot. Traditional Chinese medicine has rich experience in the treatment of gout, and the exploitation of natural XOI inhibitors shows great potential. Through deep learning and virtual screening techniques, we can rapidly screen and optimize potential active ingredients, combined with molecular docking analysis of the binding patterns of primary screening compounds, as shown in Table 5, thereby accelerating the search for natural XOI inhibitors.

Table 5. XOI training set compound names and IC50 values

| Compound number | Conpen Nongbo | IC50 (nmol) |
| --- | --- | --- |
| 1 | 2-Amino-4-(3-cyano-4-isobutoxyphenyl) | 0.59 |
| 2 | 2-(3-Cyano-1-isopropyl-1H-indole) | 0.94 |
| 3 | 4-(4-Butane-2-oxo-3-cyanobenzene) | 3.01 |
| 4 | Anthracene-2-based 4-carboxylate | 3.11 |
| 5 | 4-Hydroxy-4-(6-isocyano-4H-thiophene) | 10.01 |
| 6 | 2-(3-Cyano-5-methyl-1H-indole) | 6.41 |
| 7 | 5-Hydroxy-4-(5-cyano-2-fluorobenzene) | 4.00 |
| 8 | 3-Cyanocordycepin | 5.91 |

In this study, we used deep learning techniques to build and validate pharmacophore models to improve the efficiency of drug discovery. By using the 3D-QSAR pharma module in DS4.0 software, a pharmacophmore model was designed for a specific target, including hydrogen bond donor, hydrogen bond acceptor, hydrophobic region and other characteristic elements. These features were extracted by compounds in the training set to form multiple pharmacophore models, set the energy threshold to 10 kcal/mol, the maximum conformation number to 255, and keep other parameters default to ensure the comprehensiveness and effectiveness of the model.

The pharmacophore model was rigorously evaluated and validated. The difference between Null cost and Total cost, the deviation of predicted activity value, and the matching of characteristic elements were used to evaluate the model performance. Through these evaluations, the optimal pharmacophore model was selected. Then, using the Ligand and Profiler modules in DS4.0, the activity prediction and correlation analysis of the model and the test set were carried out to further verify the reliability and accuracy of the model. This process ensures the effective application of the final selected pharmacophore model in drug design.

Combined with compounds in TCMD database, potential active compounds were obtained by virtual screening. Using the Screen Library module in DS4.0, compounds are screened by selecting the optimal pharmacophore model and sorted according to Fit values, as shown in Table 6 and Figure 1, to identify effective natural compounds. After molecular docking analysis, the docking results were optimized and the mechanism of potential target molecules was discussed. The application of this series of methods significantly improves the efficiency and success rate of drug screening, and lays a foundation for subsequent experimental verification.

Table 6. Structure classification of natural XO inhibitor compounds

| Classify | Type of compound | Representative compounds | Quantity | Degree of fit |
| --- | --- | --- | --- | --- |
| A | Terpenes and volatile oils | Toad dihydroxycholic acid | 56 | 14.19 |
| B | Phenylpropanoids | Coperidol diangelica ester | 21 | 14.22 |
| C | Types of alkaloids | Confilii sadine | 14 | 14.16 |
| D | Flavonoids | Artemisin | 7 | 14.24 |
| E | Quinones | Comfrey furanquinone | 2 | 14.31 |
| F | Other | cis-4-decaenoic acid | 2 | 14.17 |

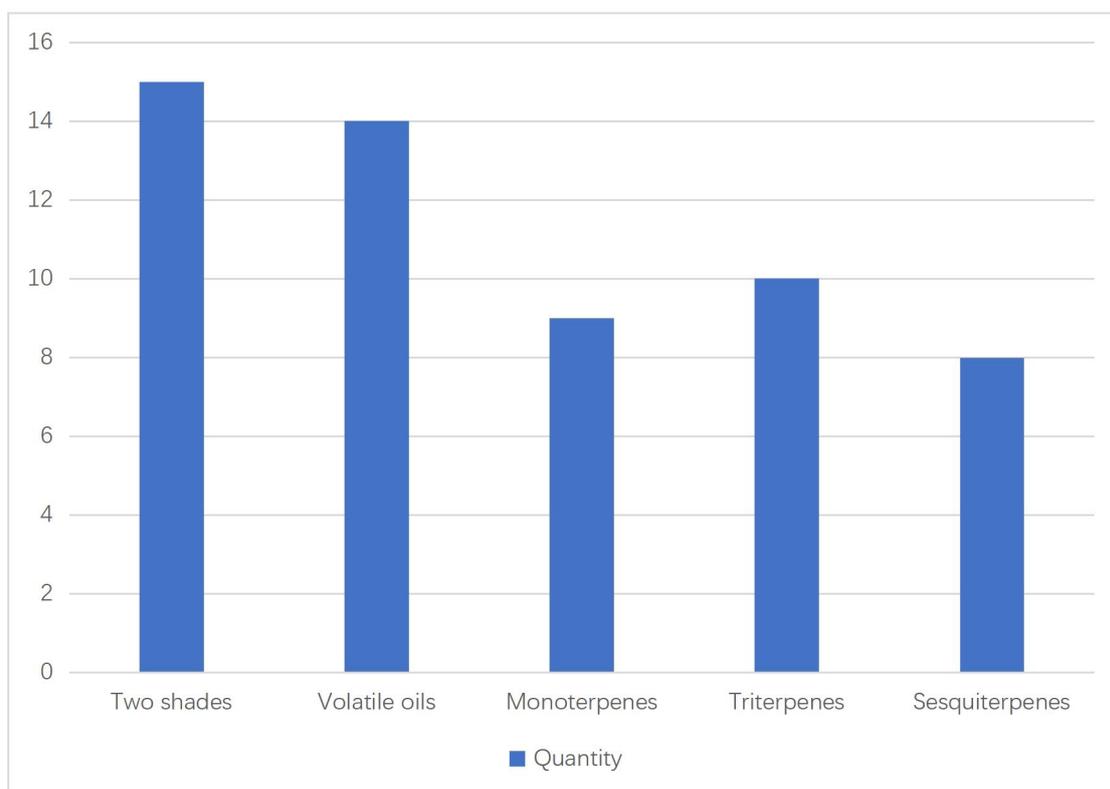

Figure 1. Classification of XO inhibitors among terpenes and volatile oils

Through the analysis of receptor-ligand interactions for the top five compounds with S-values, the results showed that the lower the S-value, the better the binding effect of the protein crystal to the active molecule. The matches of selected natural compounds with the best pharmacophore were recorded in detail and their binding patterns were demonstrated by molecular docking techniques. For example, the molecule with compound ID 3987 exhibits good binding free energy in binding to amino acid residues, demonstrating its ability to act as a potential inhibitor.

During the screening process, 102 natural XO inhibitors were identified, most of which were from chrysanthemum, legumes, and umbelliferae. By analyzing the distribution patterns of Chinese herbs, the study revealed the sources of potential XO inhibitors, highlighting the importance of certain herbs in pharmacological applications. In particular, some Chinese herbs with spicy and bitter characteristics were found to be closely related to the inhibitory activity of XO. The components and properties mentioned in the literature are also consistent with the results of this study, further supporting the feasibility of mining medicinal plants by computational methods.

Finally, the study shows that virtual screening methods using deep learning can not only quickly and efficiently identify potential inhibitors, but also guide subsequent drug development work to a certain extent. Combining with the existing theory of traditional Chinese medicine, it is pointed out that herbs mainly characterized by cold and bitter taste are more likely to play an inhibitory role. Through this method, more in-depth research can be carried out on specific ingredients in the future to find efficient and low-toxic Chinese medicine treatment schemes, so as to provide new ideas and directions for the treatment of HUA.

## 5. Conclusion and prospect

In this study, a combination of computer-aided drug design and deep learning strategies was used to systematically analyze natural product databases to identify PDE4/7 compounds with potential double inhibition. The binding stability between the PDE4/7 protein and the screening compound was evaluated by molecular dynamics simulations, thus providing scientific support for the results. At the same time, the PDENet screening model based on artificial neural network (ANN) effectively improves the screening efficiency and accuracy. After experimental verification, several compounds showed positive effects on inhibiting PDE4/7 activity and reducing inflammatory factors. This deep learning-based virtual screening method brings new ideas to the development of new drugs, especially in the discovery of compounds with medicinal potential.